\documentclass[12pt]{article}

\usepackage{amsmath,amsfonts,epsfig,color,latexsym}

\newcommand{\be}{\begin{equation}}
\newcommand{\ee}{\end{equation}}
\newcommand{\bea}{\begin{eqnarray}}
\newcommand{\eea}{\end{eqnarray}}

\renewcommand{\theequation}{\thesection.\arabic{equation}}

\let\newsection=\section
\renewcommand{\section}{\setcounter{equation}{0}\newsection}

\begin{document}

\begin{flushright}
hep-th/0501039\\
BROWN-HET-1434
\end{flushright}
\vskip.5in

\begin{center}

{\LARGE\bf The soft Pomeron from AdS-CFT }
\vskip 1in
\centerline{\Large Horatiu Nastase}
\vskip .5in

\end{center}
\centerline{\large Brown University}
\centerline{\large Providence, RI, 02912, USA}

\vskip 1in

\begin{abstract}

{\large In previous treatments, high energy QCD was analyzed using AdS-CFT
a la Polchinski-Strassler. Black hole production in AdS was responsible 
for power law behaviour of the total QCD cross section. Using the simplest 
self-consistent gravity dual assumption, that cut-off $AdS_5$ is supplemented 
by a 5d space $X_5$ of effective ``average'' 
size much larger than the scale of 
$AdS_5$, we find an energy behaviour 
just before the saturation of the Froissart bound that is $\sigma_{tot}
\sim s^{1/n}=
s^{1/11}\simeq s^{0.0909}$. It comes from the solution of the Laplacean
on $AdS_{d+1}\times X_{\bar{d}}$ behaving like $1/r^n=1/r^{2(d-1)+\bar{d}}= 
1/r^{11}$. We argue that this should be present in 
real QCD as well, as string corrections to the dual scattering are small,
and should onset at about $N_c^2 M_{1, glueball}\sim 10 GeV$. 
Experimentally, one found the ``soft Pomeron'' behaviour, 
$\sigma_{tot}\sim s^{0.093(2)}$, that onsets at about $ 9 GeV$, that 
was later 
argued to be replaced by the unitarized  Froissart + reaction-dependent  
constant 
behaviour. We argue that the soft Pomeron and the dual behaviour
 represent the same physics, creation of an effective field theory 
``soliton''-like structure (=black hole), that then decays,
 and so they have to be taken seriously. We thus have
an experimental prediction of string theory, literally counting the extra
dimensions.
}

\end{abstract}

\newpage

\section{Introduction}

At very large center of mass energies $\sqrt{s}$ (much larger than the 
hadrons mass, e.g $\gg 1 GeV$ for protons), 
the scattering of two hadrons in QCD is behaving in a ``soft'' manner: 
The total cross-section $\sigma_{tot}$ 
for the large $s$, fixed $t$ scattering has a 
slow dependence with $s$. 

But there is a bound on $\sigma_{tot}(s)$ at large energies due to 
Froissart \cite{frois}, with saturation of the type
\be
\sigma_{tot}\sim \frac{A}{M^2} ln^2 \frac{s}{s_0}
\label{froissart}
\ee
where M is the mass lightest of the lightest excitation in the 
theory and A is a constant satisfying $A\leq \pi$. In pure Yang-Mills, 
M is the mass of the lightest glueball excitation $M_{1,glueball}$, and 
if there is an almost Goldstone boson of smaller mass, like the pion of 
QCD, then $M=m_{\pi}$ and $A/M^2\leq 60 mb$. 

Experimentally, one first found the ``soft Pomeron'' behaviour \cite{compas}
(cited in the 2001 PDG \cite{pdgold}), with $\sigma_{tot}\sim s^{0.093(2)}$.
More precisely, in the scatttering of two hadrons A and B, after subtracting 
C-odd and C-even meson exchanges, one finds
\bea
&&\sigma_{AB}- Y_{1AB}(s_1/s)^{\eta_1}+Y_{2AB}(s_1/s)^{\eta_2}= 
X_{AB}(s/s_0)^{\epsilon} 
\nonumber\\&&
\sigma_{\bar{A}B}- Y_{1AB}(s_1/s)^{\eta_1}-Y_{2AB}(s_1/s)^{\eta_2}= 
X_{AB} (s/s_0)^{\epsilon}
\label{softp}
\eea
and experimentally, one finds a $\chi ^2/d.o.f.=1$ fit for energies above
$\sqrt{s}_{min}$=9 GeV (if we extend the fit to
 lower energies, $\chi^2/d.o.f.$ increases), 
giving $\epsilon= 0.0933\pm 0.0024$ and $X_{AB}\sim 10-35 mb$ \cite{compas}. 

However, later it was found \cite{compete} (cited in the 2004 PDG \cite{pdg})
that the fit is better (with $\chi^2/d.o.f =0.971$) if we replace the 
``soft Pomeron'' by the maximal Froissart behaviour, plus a constant, 
reaction-dependent term, i.e.
\bea
&&\sigma_{AB}- Y_{1AB}(s_1/s)^{\eta_1}+Y_{2AB}(s_1/s)^{\eta_2}= 
Z_{AB}+B log^2(s/s_0) 
\nonumber\\&&
\sigma_{\bar{A}B}- Y_{1AB}(s_1/s)^{\eta_1}-Y_{2AB}(s_1/s)^{\eta_2}= 
Z_{AB}+B log^2(s/s_0)
\label{froisconst}
\eea
in which case the behaviour can be extended down to $\sqrt{s}_{min}$
=5GeV, with $B=0.31 mb\ll 60 mb$ and $Z_{AB}\sim 18-65 mb$. 

Using AdS-CFT \cite{malda} for a general non-conformal theory, in \cite{kntwo}
the high energy behaviour of gauge theories was analyzed, and power law 
behaviours for $\sigma_{tot}(s)$ were found, corresponding to black hole
production in the gravity dual, that settle into the maximal Froissart 
behaviour. The saturation of the Froissart bound was proven, a fact which is 
still not done in QCD. Moreover, in \cite{knthree} the saturation behaviour 
in the gravity dual was mapped exactly onto a 1952 effective field theory 
model due to Heisenberg \cite{heis}, of collisions of shockwaves of pion 
field distributions. 

In this paper, we  will argue that one needs an extra assumption about the 
gravity dual, and then we find a power law $\sigma_{tot}\sim s^{1/11}$ 
setting in at about $N_c^2 M_{1, glueball}$, which is around 10 GeV in QCD, 
and that will settle into the maximal
Froissart behaviour. Thus we argue that the ``soft 
Pomeron'' behaviour is real, gives a string theory prediction, and one needs 
further experimental work to determine how does the argued-for 
maximal Froissart 
behaviour in \cite{compete} fit in. For previous attempts at 
describing the soft Pomeron in AdS-CFT, see \cite{janik}.

We will first explain and expand on
 the large N, large $g^2N$ analysis of high energy 
gauge theory scattering in \cite{kntwo} (section 2), then explain the 
$s^{1/11}$ behaviour and summarize the energy regimes of gauge theories
(section 3). Then we will describe what happens in 
real QCD and compare to the experimental evidence (section 4). 

\section{Using AdS-CFT duality for high energy gauge theory scattering}

Polchinski and Strassler \cite{ps} (see also \cite{pstwo}) have shown that
one can describe the scattering of colourless states 
at large energies in a gauge theory by scattering in a minimal model of 
gravity dual. A conformal theory is dual to an $AdS_5\times X_5$ space
\be
ds^2= \frac{\bar{r}^2}{R^2} d\vec{x}^2 + \frac{R^2}{\bar{r}^2}d\bar{r}^2 + 
R^2 ds_X^2= e^{-2y/R}d\vec{x}^2 + dy^2 + R^2 ds_X^2
\ee
and to describe a nonconformal theory one just cuts off the space in the IR, 
at $\bar{r}_{min}\sim R^2\Lambda_{QCD}$ (equivalently, at $y_{max}$), where 
$\Lambda_{QCD}$ is the mass of the lightest excitation of the gauge theory
(glueball). This hides our ignorance about what happens in the IR, 
corresponding to small r modifications of the gravity dual, but this simple 
model is enough to obtain many features of the gauge theory scattering. 
This cut off is equivalent to putting an IR brane, thus getting the 
Randall-Sundrum 
model \cite{rs} (if we put an optional UV cut off).

A gauge theory mode with momentum $p$ and wavefunction $e^{ipx}$ corresponds 
in the gravity dual to a mode with local AdS momentum $\tilde{p}_{\mu}
= (R/\bar{r}) p_{\mu}$ and wavefunction $e^{i\tilde{p}x}\psi(\bar{r}, 
\Omega)$, and the string tension $\alpha' = R^2/(g_sN)^{1/2}$ 
corresponds to the gauge theory 
string tension $\hat{\alpha}' =\Lambda_{QCD}^{-2}/(g^2_{YM}N)^{1/2}$
and $\sqrt{\alpha '}\tilde{p}_{string}\leq \sqrt{\hat{\alpha}'}q_{QCD}$. 
The gauge theory amplitudes are related to string amplitudes by 
\be
{\cal A}_{gauge}(p)= \int d\bar{r}
 d^5 \Omega \sqrt{g} {\cal A}_{string}(\tilde{p})\prod_i \psi_i
\label{polstra}
\ee
At large $\bar{r}$, $\psi$ behaves as 
\be
\psi (\bar{r},\Omega)\sim C (\bar{r}/\bar{r}_{min})^{-\Delta}g(\Omega)
\ee
Since ${\cal A}_{string}= {\cal A}_{string}(\tilde{s}, \tilde{t})$, 
one takes $\nu =-\alpha ' \tilde{t}$ as integration variable (then 
$\bar{r}= \nu ^{-1/2} \bar{r}_{min}$ $ \sqrt{\hat{\alpha}'t}$), with 
${\cal A}_{string}= {\cal A}_{string}(\nu s/|t|, \nu)$. 
If one takes the large $\bar{r}$ behaviour of the wavefunctions to be valid 
everywhere, one finds that 
most of the integration in the high energy ($s\rightarrow \infty$) 
case is situated in the IR
(small $\bar{r}$). However, as we can easily see, the fact that the 
wavefunction 
gets modified at small $\bar{r}$ 
will only modify the behaviour of ${\cal A}_{gauge}$
with t, the s behaviour still comes from ${\cal A}_{string}$. What can 
also happen is that the modification of the wavefunction is such as to 
keep the bulk of the integral centered not on $\bar{r}_{min}$, 
but a finite distance away from it. 

Giddings then noticed that one will produce black holes in the gravity 
dual when one reaches the Planck scale $M_P= g_s^{-1/4}{\alpha '} ^{-1/2}= 
N^{1/4} R^{-1}$,
corresponding to the gauge theory scale $\hat{M}_P= N^{1/4}\Lambda_{QCD}$
\cite{gid}. The black hole horizon radius in flat D dimensions grows with mass 
like $r_H\sim M^{1/(D-3)}$, thus if the cross section for black hole 
formation is assumed to be a simple black disk with radius $r_H(M=\sqrt{s})$,
the cross section for black hole formation will grow like
\be
\sigma \sim \pi r_H^2\sim (\sqrt{s})^{\frac{2}{D-3}}
\ee
For D=5 (only AdS is approximately flat), $\sigma \sim s^{1/2}$, whereas 
if D=10 ($AdS_5\times S_5$ is approximately flat), $\sigma \sim s^{1/7}\simeq
s^{0.143}$. 

As the black holes grow in size, the horizon of the 10d black hole will 
reach the AdS size when $E=E_R= M_P (RM_P)^7= N^2 R^{-1}$, 
corresponding in gauge theory to  $\hat{E}_R= N^2 \Lambda_{QCD}$, and this 
was argued that should correspond to the maximal 
Froissart behaviour in gauge theory. 
Indeed, if one takes the linearized gravity induced by a point mass $m=\sqrt{s}
$ on the IR brane, and obtains the horizon radius for it by setting the 
perturbation to 1, one obtains
\be
h_{00, lin} \sim G_4 \sqrt{s} \frac{e^{-M_1r}}{r}\sim 1\Rightarrow 
\sigma\sim \pi 
r_H^2 \sim \frac{\pi}{M_1^2} ln^2 (\sqrt{s}G_4 M_1)
\ee
which with the assumption $\sigma_{QCD}\sim \sigma$ is just the 
maximal Froissart 
behaviour, in the case the mass M is the lightest glueball mass, corresponding 
to the lightest KK graviton in the dual, $M_1= j_{1,1}/R$. Then indeed, 
$r_H\sim 1/M_1\sim R$. 

The case when the pion (almost Goldstone boson) is lightest is treated 
similarly, by making the radion of the Randall Sundrum model dynamical, 
with mass $M_L$. Then the IR brane bends under the mass, and now  
\be
\frac{\delta L}{L}|_{lin} \sim G_4 \sqrt{s} (M_L R) \frac{e^{-M_1r}}{r}
\sim 1\Rightarrow \sigma_{QCD}= \sigma \sim 
\frac{\pi}{M_L^2}ln^2 (\sqrt{s}G_4 M_L R)
\label{goldstone}
\ee
and one gets the Froissart bound with $M_L\leftrightarrow m_{\pi}$. 

This simple analysis was made more rigorous and exact 
(i.e. calculating coefficients) in \cite{kntwo} and 
was mapped exactly to the Heisenberg model \cite{heis} for the saturation 
in \cite{knthree}. This was done as follows.

The scattering at high energies in the gravity dual can be described by 
scattering of Aichelburg-Sexl shockwaves \cite{as} in the gravity dual. 
In flat D dimensions, the shockwaves are
 \be
ds^2 = 2dx^+ dx^- +(dx^+)^2 \Phi (x^i) \delta (x^+) +d\vec{x}^2
\ee
where the function $\Phi $ satisfies the Poisson equation
\be
\Delta_{D-2} \Phi (x^i) = -16 \pi G p \delta ^{D-2}(x^i)
\label{poisson}
\ee
One can put these shockwaves inside gravity duals \cite{nastase}, and 
one takes advantage of the fact that one still has the Poisson equation 
(\ref{poisson}) for the function $\Phi$, just the Laplacean is taken in 
the background. Thus for the shockwaves, the linearized solution is the 
exact solution, and one can find it explicitly, unlike the black hole 
solution in the background.  For early shockwave-type solutions in AdS
and AdS-CFT see \cite{hi}, used to argue for black hole creation \cite{bf}.
The linearization of shockwaves in AdS and on braneworlds was observed 
in \cite{cg}.

At energies below the Planck scale $M_P$ (in gauge theory 
$\hat{M}_P=N^{1/4}\Lambda_{QCD}$) , or rather below the string scale
$E_s= {\alpha '}^{-1/2}$ (in gauge theory $\hat{E}_s= (g^2_{YM}N)^{1/4}\Lambda
_{QCD}$), one takes only one of the scattered 
particles as a shockwave, and the second as a null geodesic scattering in 
the background \cite{thooft}. One can calculate this 't Hooft scattering 
in gravity duals \cite{nastase}, and in \cite{kntwo} the behaviour of 
the AdS amplitude ${\cal A}_{string}\sim G_4 s$ was found, giving 
\be
\frac{d \sigma_{AdS}}{d^2 k}= \frac{4}{s} \frac{d\sigma_{AdS}}{d
\Omega} \sim (G_4 s)^2
\ee
and correspondingly in gauge theory
\be
{\cal A}_{gauge} \sim \hat{G}_4 s\Rightarrow 
\frac{d \sigma_{gauge}}{d^2 k}= \frac{4}{s} \frac{d\sigma_{gauge}}{d
\Omega} \sim (\hat{G}_4 s)^2
\ee

At energies above the string scale $E_s= {\alpha '}^{-1/2}$ 
(in gauge theory $\hat{E}_s= (g^2_{YM}N)^{1/4}\Lambda_{QCD}$), one has 
Regge behaviour of the string amplitude, which implies Regge behaviour for 
the gauge amplitude (from (\ref{polstra}) 
\be
{\cal A}_{gauge}^{2\rightarrow 2} \sim (\hat{\alpha}' s)^{2+\hat{\alpha}' t
/2}
\ee

At energies above the Planck scale $M_P$, 
one has to take both particles that scatter
as shockwaves. The metric in the interacting region can only be calculated 
perturbatively away from the interaction point \cite{dp}, but luckily  
one can calculate the presence of a ``trapped surface'' at the interaction 
point, and by a GR theorem there will be a horizon forming outside it, 
away from the interaction. One can calculate the trapped surface at nonzero 
impact parameter b, and derive a maximum $b_{max}(s)$ for which a trapped 
surface forms. This formalism was put forward in flat 4d in \cite{eg} and 
generalized to curved higher d in \cite{kn}, with an approximation scheme 
for $b_{max}(s)$. 

Once we have a $b_{max}(s)$ describing this classical scattering, one can 
use a simple eikonal model to get a quantum amplitude from it, 
with the eikonal being the simplest thing we can have, a black disk
\be
Re(\delta(b,s) )=0\;\;\; Im (\delta(b,s) )=0,\;b>b_{max}(s);\;\;\;
Im (\delta(b,s) )=\infty , \; b<b_{max}(s)
\ee
Then the resulting quantum amplitude can be put in the Polchinski-Strassler 
formula (\ref{polstra}) to derive a gauge amplitude, 
and the result is that the classical gravity $\sigma_{tot}$ gets multiplied  
by a model-dependent constant (depends on the details of the gravity dual, 
that we have approximated by the RS model), and we replace $R$ by $\Lambda
_{QCD}$ and $M_P$ by $N^{1/4}\Lambda_{QCD}$ (gravity parameters replaced 
by gauge parameters). 

This scattering model looks very much like Heisenberg's model \cite{heis}, 
with an exact match for the Froissart saturation, as shown in \cite{knthree}.
Heisenberg argues that at high energies, the hadrons scattering are replaced
by pion field distributions that look like shockwaves, because of very large 
(infinite) Lorentz boosts. While Heisenberg had pion fields, in the case 
we are describing, of only Yang-Mills, we have instead 
the lightest glueball field, mapped to gravity excitations in the dual. 
Indeed, in the gravity dual description, we have collisions of gravity 
field shockwaves. 

The nonlinearity of the pion field, described by Heisenberg through the 
DBI-like action
\be
S= l^{-4}\int d^4x \sqrt{1+l^4[(\partial_{\mu} \phi)^2 +m^2\phi^2]}
\label{dbi}
\ee
and in our case by the nonlinearity of the gravitational action in the 
dual (and of the glueball field in the gauge theory), is responsible 
for creating a nonlinear ``soliton'' in the collision. One cannot 
find it in perturbation theory (Heisenberg presents the perturbative 
pion solution), as seen also in our case: one can't find the black hole in 
perturbation theory for the A-S collision \cite{dp}. The soliton will decay
through emission of pions in Heisenberg's model, and emission of 
gravitons for the black hole gravity dual.

For Heisenberg's model, one has at $x^+\leq 0, x^-\leq 0$, 
\be
\phi= \phi_1+\phi_2= \delta (x^+) \psi_1 (x^i) + \delta (x^-)\psi_2 
(x^i)
\ee
where $x^i$ are transverse coordinates, and the evolution of this shockwave 
should give the ``soliton'' at $x^+>0, x^->0$. 

What brings in the saturation of the Froissart bound is the assumption that 
$\psi (x^i)\sim e^{-m_{\pi} r}$ and the ``degree of inelasticity'' $\alpha$
(=${\cal E}/ \sqrt{s}$= energy loss/collision energy) behaves 
similarly to $\psi$ 
as a function of the impact parameter b (see also \cite{knthree}
for a more detailed account).  

But if we don't make this extra assumption, we get an effective field theory 
model in the gauge theory for all black hole formation in the gravity dual.

Indeed, in the gravity dual, we scatter two A-S shockwaves, and for $
M_P<\sqrt{s} < E_R$ (corresponding in gauge theory to $\hat{M}_P= N^{1/4}
\Lambda_{QCD}<\sqrt{s}< \hat{E}_R= N^2\Lambda_{QCD}$), 
the black holes formed can be considered to be 
in flat space. Correspondingly, we take A-S shockwaves in flat space, 
thus solutions to (\ref{poisson}) in flat D-dimensional space, for which 
($r=\sqrt{x^ix^i}$)
\be
\Phi = \frac{16\pi G_D}{\Omega_{D-3}(D-4) r^{D-4}}\sim \frac{1}{r^{D-4}}
\ee
Then one obtains the maximum b for black hole formation \cite{kntwo}
\bea
&&b^2_{max}\leq 2[\frac{\alpha}{D-2}]^{\frac{1}{D-3}}\frac{D-3}{D-2}= 
(D-3)\frac{2(\epsilon r_H)^2}{ [D-2]^{\frac{D-2}{D-3}}}\nonumber\\
&&r_H=[\frac{16\pi G_D \sqrt{s}}{(D-2)\Omega_{D-2}}]^{\frac{1}{D-3}}
\;\;\;\epsilon=[\frac{(D-2)\Omega_{D-2}}{4\Omega_{D-3}}
]^{\frac{1}{D-3}}
\eea
and therefore $b_{max}(s)\simeq a s^{\frac{1}{2(D-3)}}$. Then using 
(\ref{polstra}) the gauge theory amplitude is 
\be
\sigma_{gauge}= \bar{K}\pi a^2 (\frac{\hat{\alpha}'s}{\alpha '})
^{\frac{1}{D-3}}
\ee
where $\bar{K}$ is a model dependent constant and in $a$ we need to replace 
$G_D$ by $\hat{G}_D$ (gauge theory quantity). Again, for D=5 we get 
$\sigma \sim s^{1/2}$, and for D=10 we have $\sigma \sim s^{1/7}\simeq 
s^{0.143}$, but now we have an exact picture. 

Now we see that we can also map to Heisenberg's model, if we only relax 
the assumption about the form of $\psi(x^i)$ (which was natural for a 
pion of mass $m_{\pi}$), and say that $\psi (x^i)$ is instead mapped to 
$\Phi\sim 1/r^{D-4}$. Of course, the caveat is that the gravity dual 
picture is intrinsically D-dimensional, whereas the Heisenberg model is 
in 4d, but this is just the usual holography. 

For the maximal Froissart behaviour however, the Heisenberg and dual 
descriptions match exactly. When the horizon size of the formed black holes
becomes comparable with the AdS size, namely at $E=E_R$ in AdS and $\hat{E}_R
= N^2\Lambda_{QCD}$ is gauge theory, we have to consider the curvature of 
space into account. But as we saw, most of the integral in the 
Polchinski-Strassler formula (\ref{polstra}) is situated at the IR end (IR
brane), if the large $\bar{r}$ behaviour of the wavefunctions remains 
(or is not 
modified too much). In that case, the black holes being mostly created near 
the IR brane will eventually be large enough not to feel they are away 
from the IR brane. 

One takes then the scattering of two A-S shockwaves on the IR brane, that 
behave at large radius as 
\be
\Phi(r, y=0)\simeq R_s \sqrt{\frac{2\pi R}{r}}C_1 e^{-M_1 r}; \;\;
C_1=\frac{j_{1,1}^{-1/2}J_2(j_{1,1})}{a_{1,1}};\;\; J_1(z)
\sim a_{1,1}(z-j_{1,1});\;\;z\rightarrow j_{1,1}
\label{kkwvfct}
\ee
and one sees the same behaviour as one had for the static black hole 
perturbation $h_{00, lin.}$, namely the exponential drop, just the power of 
r and constants are different. This shockwave is the solution of the laplacean 
for a massless particle on the IR brane, and now satisfies exactly Heisenberg's
description, and is 
also a 4d picture (the higher dimensional gravity was in 
a sense KK reduced for this solution, which lives on the IR brane. The 
parameter $M_1=j_{1,1}/R$ is the first KK mass). 

From the scattering of two such waves one finds the maximum impact parameter 
that forms a black hole, 
\be
b_{max}(s)= \frac{\sqrt{2}}{M_1}ln [R_s M_1 K];\;\;\; K= \frac{3\sqrt{\pi}}{
\sqrt{2}j_{1,1}^{3/2}}\simeq 0.501
\ee
where $R_s= G_4\sqrt{s}, G_4= 1/(RM_{P,5}^3), M_1= j_{1,1}/R$ ($j_{1,1}\simeq 
3.83$) and the gauge theory cross section is 
\be
\sigma_{gauge}= \bar{K}\pi b_{max}^2(\tilde{s})
\ee
where as before we must replace gravity quantities with gauge theory 
quantities, and $\bar{K}$ is a model dependent constant.

In \cite{kntwo} a possible intermediate case was analyzed also, when 
the black holes that are formed start feeling the AdS size, but not the 
IR brane yet. In that case, at large r, an A-S shockwave inside $AdS_5$ 
was found to behave like 
\be
\Phi=\frac{\bar{C}R^4}{r^6}e^{(4y+2y_0)/R}
\label{adsphi}
\ee
where $\bar{C}= 2R_s R^2$ and the shockwave lives at at $y=y_0$ in $AdS_5$.
Then in the scattering of two such waves, one finds the maximum b for 
black hole formation, and the corresponding scattering cross section
\be
b_{max}=7^{-1/12}\sqrt{\frac{12}{7}}R e^{y_0/R}[\frac{R_s}{Re^{y_0/R}}]^{1/6}
\Rightarrow \sigma_{gauge}=\pi {a'}^2\bar{K}(\frac{\hat{\alpha}'s}{\alpha '})
^{\frac{1}{6}}
\label{adsimpact}
\ee
thus $\sigma_{gauge}
 \sim s^{1/6}$, but it was not clear whether there exists an 
energy regime corresponding to this behaviour, or whether at $E_R$ the 
gauge theory goes directly into maximal Froissart behaviour.

\section{The ``soft Pomeron'' in gauge theories; overview of energy 
regimes}

Let us analyze this problem in more detail. We have seen that creating 
black holes in flat D dimensions generates a gauge theory behaviour 
$\sigma _{gauge}\sim s^{1/(D-3)}$, specifically $s^{1/2}$ for D=5 and
$s^{1/7}$ for D=10. Since it is normal to take D=10, it would be very unlikely 
to have $s^{1/6}$, corresponding to black hole creation in $AdS_5$,
 as an intermediate behaviour  before going over to the $ln^2 s$
Froissart behaviour. On the other hand, $AdS_5$ implies D=5, and then 
we would have 
$s^{1/2}\rightarrow s^{1/6}\rightarrow ln^2 s$ which 
seems OK. But why would then 
the compact space $X_5$ be so small as not to be felt at all by the black 
holes that feel $AdS_5$ as flat? If however D=10, it seems that 
the ``black holes in $AdS_5$'' regime is excluded, so there would be no 
problem.

But let us think more about what this means. We have seen that most of the 
gauge theory amplitude (\ref{polstra}) at large s, fixed t is situated 
in the IR, near the IR brane, at least if the large $\bar{r}$ behaviour of the 
wavefunctions is not modified too drastically. But if one can first 
consider scattering in flat space as a good approximation (as opposed to 
scattering on the IR brane from the begining), first as single graviton 
exchange ${\cal A}_{gauge}\sim (\hat{G} s)^2$, then as Regge behaviour
${\cal A}_{gauge}^{2\rightarrow 2} \sim (\hat{\alpha}' s)^{2+\hat{\alpha}' t
/2}$, then as black hole creation in flat space, that means the scattering 
should first feel the curvature of AdS, and just after that to feel only 
the IR brane. 

On the other hand we have seen that black hole creation in just $AdS_5$ 
will not do, we get a 
behaviour $s^{1/7}\rightarrow s^{1/6}\rightarrow ln^2 s$ that is hard to 
imagine in the gauge theory. If the compact space $X_5$ is of size 
comparable to $AdS_5$, when $r\gg R$ the compact space will not be felt, 
and we get the same inconsistent result.

What about if $X_5$ has a much larger size? We could say that the 
wavefunctions $\psi(\bar{r}, 
\Omega)$ in the IR are such that the average position $y_{av}$ 
(where most of the AdS scattering takes place in the Polchinski-Strassler 
formula) is far from the IR brane, and we have a large average size 
$e^{2y_{av}/R}\bar{R}^2
d\Omega_5^2$ of the compact space. But this will not do, as 
we can see from (\ref{adsphi}) and (\ref{adsimpact}): the effective scale 
of AdS in the 4d theory is actually $Re^{y_0/R}$, and that is compared 
to $\bar{R}e^{y_0/R}$, where $\bar{R}$ is the scale of the compact space.

So we need the effective size $\bar{R}$ of the compact space to be much 
larger than R. That is possible, and would even solve the problem of 
having most of the scattering in (\ref{polstra}) away from the IR brane.
Indeed, the AdS wavefunctions $\psi(\bar{r})$ are modified at small $\bar{r}$, 
but on top
of that, AdS space itself will be modified at small $\bar{r}$, which can be 
modelled by $\bar{R}=\bar{R}(\bar{r})$. Then  as we can see from 
(\ref{polstra}), if $\sqrt{g_X}=\bar{R}(\bar{r})^5$ 
(the volume of the compact space) increases 
sufficiently with $\bar{r}$, 
it will balance the $\bar{r}^{-\beta}$ behaviour of the 
integral, which drives the bulk of it towards $\bar{r}_{min}$. 

Thus this simple model, that in the IR the effective size $\bar{R}$ of the 
compact space increases with $\bar{r}$, can make the average $<\bar{r}>$ 
at which most of 
the integral is situated to be $\gg \bar{r}_{min}$, 
and correspondingly also the 
effective average size $<\bar{R}>$ of the compact space to be $\gg R$.
Does this fix our gauge theory contradiction in $\sigma_{tot}(s):
s^{1/7}\rightarrow (s^{1/6}?)\rightarrow ln^2 s$?

We have analyzed this question in detail in the Appendix. In the case that 
$<\bar{R}>\gg R$, 
we have an intermediate regime where the compact space can be 
considered as approximately flat. We have solved the Poisson equation to 
get the A-S shockwave in $AdS_{d+1}\times X_{\bar{d}}$ in that case, 
(\ref{asbgr}). At large 4d distances r and $y=y_0$ ($y_0$=
interaction point), it behaves like
\be
\Phi=\frac{K_1 R_s R^n}{r^n}\sim \frac{1}{r^{2(d-1)+\bar{d}}}= \frac{1}{r^{11}}
\ee
Specifically, for d=4, $\bar{d}=5$, we have
\be
\Phi (r\gg R, y=0)= \frac{945}{8}\frac{R_s}{(2\pi)^2}\frac{R^{11}}{r^{11}}
\ee

One then scatters two of these shockwaves and calculates the trapped 
surface formed in the collision, using the formalism in \cite{kn,kntwo}.
One finds a trapped surface that satisfies, at nonzero impact parameter b,
\be
(\frac{3\Phi|_{y=0}}{2R})^2(1-\frac{b^2}{2r^2})=1
\ee
and a trapped surface that is there in the absence of the highly curved 
AdS, and is smaller. One takes the larger trapped surface as describing 
best the horizon of the black hole formed in the collision, and gets 
\be
b_{max}= \sqrt{\frac{2n}{n+1}}(n+1)^{-1/n}R (\frac{3K_1R_s}{2R})
^{\frac{1}{n}};\;\; R_s=G_4 \sqrt{s}
\ee
thus a cross section $\sigma=\pi b_{max}^2$ and a QCD cross-section 
which contains a model dependent multiplicative constant, and 
converts gravity quantities to gauge quantities ($G_4\rightarrow \hat{G}_4,
R\rightarrow \Lambda_{QCD}$).
\be
\sigma_{gauge}=\bar{K}\pi b_{max}^2(\hat{\alpha}'s/\alpha ')^{1/n}
\sim s^{1/n}
\ee
with n=11 for $d=4,\bar{d}=5$.

We see therefore that now we have indeed solved the gauge theory contradiction
for the $\sigma_{gauge}(s)$ flattening behaviour. Now we have 
$s^{1/7}\rightarrow s^{1/11}\rightarrow ln^2 s$, that is consistent flattening.

So let us review the energy regimes of gauge theories. The gauge theory energy 
scales are, in increasing order. First, the AdS scale (1/R), corresponding
to $\hat{E}_{AdS}=\Lambda_{QCD}$. Then, the string scale, corresponding to 
$\hat{E}_S ={\hat{\alpha}}'{}^{-1/2}=\Lambda_{QCD}(g^2_{YM}N)^{1/4}$. 
After that, one reaches the Planck scale, corresponding to $\hat{M}_P=
N^{1/4}\Lambda_{QCD}$, followed by the correspondence principle scale, 
at which the string description is replaced by black hole description, 
corresponding to $\hat{E}_c= \Lambda_{QCD}N^2/(g^2_{YM}N)^{7/4}$. 
Finally, one reaches the scale at which the black hole horizon size 
equals the AdS size, corresponding to $\hat{E}_R=N^2\Lambda_{QCD}$.
 
If the gravity dual would be intrinsically 5d (such that the compact space 
is always much smaller), one would have two further scales. The scale 
at which $G_4s\sim 1$ corresponds in gauge theory to $\hat{M}_{P,4}
=N^{3/8}\Lambda_{QCD}$, and the scale at which $R_s=G_4\sqrt{s}$ is 
of the order of $R_{AdS}$, corresponding in gauge theory to $\hat{E}_R'
=N^{3/4}\Lambda_{QCD}$, which is thus $\hat{E}_R$ for pure $AdS_5$ gravity 
dual. As we said, the possibility that the dual is always 5d (and the 
compact space always very small) seems hard to imagine, but is a 
self-consistent one, so we mentioned it anyway. 

Pictorially, one has the energy regimes 0, I, II, III, IV, V:
\be
0\rightarrow | \Lambda_{QCD} \;\;\; I\rightarrow |\hat{E}_s\;\;\; II 
\rightarrow |\hat{M}_P\;\;\; III\rightarrow (\hat{M}_{P,4};\;\; \hat{E}_c;
\;\;\; \hat{E}_R')\rightarrow |\hat{E}_R\;\;\; IV \rightarrow ?\;\;\;V?
\ee
and we put a question mark because there could be one more scale involved.

Let us explain what happens in each regime. In the regime I, above $\Lambda
_{QCD}$ and before $\hat{E}_s$, 
but close to it, in the gravity dual we have single graviton 
exchange, described by 't Hooft scattering: one particle creates a shockwave,
the other moves on a null geodesic. Actually, for this 
behaviour to be isolated from other behaviours, we need $\hat{M}_P$ close to 
$\hat{E}_s$, since as 't Hooft showed, we need actually 
to have energies
close to the Planck scale, not the string scale. We also need to 
be far away from $\Lambda_{QCD}$ (1/R in gravity), so that we don't feel 
the glueball masses (don't feel the AdS curvature in gravity). This is a 
stringent constraint on the gauge theory, but it could be satisfied in 
principle.
In gauge theory this would also correspond
to exchange of a single universal colourless ``graviton'', which would be 
a nonrenormalized version of the ``Pomeron'', with intercept $\alpha(0)=2$ 
(graviton). 

Further in energy, in regime II, we have Regge behaviour for the 
string amplitude, and correspondingly Regge behaviour for the gauge theory
\be
{\cal A}_{gauge}^{2\rightarrow2}\sim (\hat{\alpha}'s)^{2+\hat{\alpha}'t/2}
\ee
Now we have Regge trajectories $\alpha(t)= 2+\hat{\alpha}'t/2$ replacing 
the ``graviton'' exchange.

In regime III, above the Planck scale, we will start producing black holes, 
corresponding in the gauge theory to nonlinear solitons of the glueball 
effective field (via the Heisenberg description, extended in this paper to
non-saturated behaviour). We will not treat the case of the pure 5d gravity 
dual. Then, if the dual is 10d, 
we get $\sigma_{gauge}(s)\sim s^{1/(D_{tot}-3)}=s^{1/7}$, from the 
decay of the glueball effective field soliton=black hole. 

The black holes being created we have argued that happen 
on the average at a $<\bar{r}> \gg \bar{r}_{min}$ 
for consistency, and thus one doesn't
feel the IR brane yet. One first reaches AdS size, 
when $E= \hat{E}_R$, entering regime IV, and then in 
$AdS_{d+1}\times X_{\bar{d}}$ one gets 
\be
\sigma_{gauge}\sim s^{\frac{1}{n}}= s^{\frac{1}{2(d-1)+\bar{d}}}=s^{\frac{1}{
11}}
\ee

As the black holes continue to grow, they will eventually reach the 
IR brane and grow as large as to be effectively on the IR brane. That should
happen at an unknown scale $\hat{E}_F$, depending on $M_1$ (the mass of 
the lightest glueball, and the mass of the lightest KK graviton in the 
gravity dual), 
which would signal the onset of 
the maximal Froissart behaviour in an energy regime V. 
This scale would depend on the details of the gravity dual.

Up to now we have analyzed the case that the lightest excitation is a glueball.
But if the lightest excitation is an almost Goldstone boson like the pion 
of QCD, the maximal Froissart behaviour will be in terms of the pion field.
The simple order-of-magnitude argument of \cite{gid} showed that if the 
pion is the radion in the RS model, a similar picture emerges, with the 
IR brane bending under the mass $\sqrt{s}$. It is not clear how to translate 
this into a picture involving collision of waves, but it is clear that somehow 
the bending will become larger first, engulfing the black hole in it. 
Therefore one will have another unknown scale $\hat{E}_F'$ which will be 
$\hat{E}_F (M_1\rightarrow M_L)$ (replace the first glueball mass by the 
pion mass, or KK graviton mass by the radion mass in the gravity dual), 
and $\hat{E}_F'<\hat{E}_F$. 
At that scale, one will have maximal Froissart behaviour in terms of 
$M_L=m_{\pi}$.

\section{Real QCD and experiments}

So what should happen in real QCD? First, one needs an argument that all 
that we said still applies in QCD. In \cite{kn}, string corrections to the 
black hole production via A-S scattering were computed in flat d=4, and 
they were used in \cite{kntwo,knthree}. One scattered string-corrected 
A-S shockwaves and analyzed their effect on the black hole production. 
The string corrected shockwaves, due to Amati and Klimcik \cite{ak} are 
obtained as follows. One matches the 't Hooft scattering of a superstring 
in an arbitrary shockwave profile $\Phi$, given by $S=e^{ip^+\Phi}$, with 
a resummed eikonal superstring calculation \cite{acv87}, $S=e^{i\delta}
=\sum_h(i\delta)^h/h!\sim \sum_h (g_s)^h (a_{tree})^h/h!$ (h= loop number). 
By equating the two results, one gets
\be
\Phi(y)= -q^v \int_0^{\pi}\frac{4}{s} : a_{tree} (s, y- X^d (\sigma_d, 0))
:\frac{d\sigma_d}{\pi}
\ee
where $b\equiv x^u-x^d$ is the impact parameter, that becomes the variable y.
This procedure contains both $\alpha'$ and $g_s$ corrections. The $g_s$ 
corrections come from the eikonal resummation, obtained by gluing tree 
amplitudes into ``ladder diagrams''. The $\alpha'$ corrections come from 
the fact that $\Phi$ at large y is Aichelburg-Sexl + $\alpha '$ corrections 
(from $a_{tree}$). When scattering two string-corrected AK shockwaves, 
one obtains a maximum impact parameter \cite{kn}
\be
B_{max}= \frac{R_s}{\sqrt{s}}(1+ e^{-\frac{R_s^2}{8\alpha ' log (\alpha ' s)}}
)
\label{corrections}
\ee
when the exponent is large in absolute value, and the uncorrected term is 
the A-S result. The condition for the exponent to be large, when one 
dimensionally reduces to 4d, gives 
\be
G_4 s \frac{G_4/\alpha '}{log (\alpha 's)}>1
\ee
or $\sqrt{s}>E_0$, and replacing the formulas that we have in our case gives a
QCD energy scale
\be
E_0\sim \frac{\hat{M}_P^2}{\hat{E}_s}\sim\Lambda_{QCD}N^{1/4} g_{YM}^{-1/4}
\ee
As the scattering was not done in our case, we can't trust $E_0$, but in 
any case we see that it is most likely smaller than $\hat{E}_R$, the scale 
of the onset of the ``soft Pomeron'' behaviour $\sigma_{QCD} \sim s^{1/n}$. 
Why are the calculated
string corrections exponentially small? We have no good 
physical argument for 
it, other than the argument 't Hooft gives for the predominance of 
't Hooft scattering over any massive interaction corrections: massive 
interactions are finite range, and they get infinitely time delayed due 
to the divergence of $\Phi$ at small r. But this argument, extended to the 
black hole production case, is not completely similar, since for instance 
it was also found in \cite{kn} that when the exponent in (\ref{corrections})
is small in absolute value, we get very large (positive, i.e.
increasing $B_{max}$) corrections. 

In the gauge theory, string $\alpha'$ and $g_s$ corrections translate 
into $1/N$ and $1/(g_{YM}^2N)$ corrections, thus if string corrections 
are small, we can apply our calculations in real QCD ($N=N_c=3$ and 
$g_{YM}\sim 1$) as well.

So we know that string corrections to the scattering will be insignificant
at $s\rightarrow \infty$, and most likely also above $\hat{E}_R$. But there
will be of course corrections to the gravity dual itself. However, as we have 
not used any details of the gravity dual other than the scales, we know  
that at most, we can have renormalization of the energy scales, e.g.
$R^{-1}\leftrightarrow \Lambda_{QCD}$,
$M_P\leftrightarrow \hat{M}_P= N^{1/4}\Lambda_{QCD}$
 and $E_R\leftrightarrow \hat{E}_R=
N_c^2 \Lambda_{QCD}$.

In QCD, the relevant energy scales are as follows. We have the pion mass 
$m_{\pi}\simeq 140 MeV$ that will appear in the Froissart bound 
(\ref{froissart}). The mass of the lightest glueball is not known, as it 
was not discovered yet. On the lattice, one finds $M_1\sim 1.6 GeV$ 
\cite{pdgtwo}, and experimental candidates range from $0.6GeV$ to $1.7 
GeV$. Overall, $\Lambda_{QCD}\equiv M_1\sim 1GeV$, thus $\hat{E}_R
=N_c^2 \Lambda_{QCD}\sim 10GeV$. We cannot be exact here, as factors of 
two, as well as the renormalization of energy scales could modify this 
result. Then also $\hat{M}_P=N_c^{1/4}\Lambda_{QCD}\simeq 
\hat{E}_s=(g^2_{YM}N)^{1/4}\Lambda_{QCD}\simeq \Lambda_{QCD}$. 

So what do we expect to happen from the analysis in the previous section? 
The regimes I and II are practically nonexistent. However, Regge behaviour 
will be present in the elastic ($2\rightarrow 2$) part of the amplitude, even 
in the region III, and is indeed observed. 

In region III, that is from about $\hat{M}_P\sim
1-2GeV$ to about $\hat{E}_R\sim 10GeV$, we would expect 
to create black holes in flat space in the gravity dual, with $\sigma_{QCD,tot}
\sim s^{1/7}\simeq s^{0.143}$, but it is not clear that the energy 
regime is large enough. Also, as we have seen, in this region we could still
maybe have large string corrections to the scattering, so our analysis 
is not guaranteed here. 

In region IV, above $\hat{E}_R\sim 10 GeV$, we expect to go over to the 
``soft Pomeron'' behaviour, where we create black holes in $AdS_{d+1}\times 
X_{\bar{d}}$, and get $\sigma_{QCD,tot}\sim s^{1/n}= s^{1/[2(d-1)+\bar{d}]}
=s^{1/11}$. 

Since in QCD $m_{\pi}\simeq 140 MeV< M_1\sim 1 GeV$, the 
maximal Froissart behaviour will be in terms of $m_{\pi}$, dual 
to something like the radion mass $M_L$. It should onset at an unknown 
energy scale $\hat{E}_F'(m_{\pi})$, depending on the details of the 
gravity dual. 

On the experimental side, as we said in the introduction, in \cite{compas}
(cited in \cite{pdgold}), one found the ``soft Pomeron'' behaviour
(\ref{softp}), 
where $\sigma_{tot}\sim s^{0.093(2)}$, more precisely $\sigma_{tot}\sim 
s^{\epsilon}$, with $\epsilon=0.0933\pm 0.0024$, and 
this behaviour fits well all data above $\sqrt{s}_{min}= 9GeV$, with a 
$\chi^2/d.o.f.=1$ (if one tries to extend this behaviour to lower energies, 
$\chi^2$ increases).  Thus we have remarkable agreement with our predicted 
soft Pomeron behaviour! 

Strangely though, later it was found \cite{compete} (cited in \cite{pdg})
that a statistically better fit (with $\chi^2/d.o.f.=0.971$) is given by
a maximal Froissart behaviour, plus a reaction dependent constant term
(\ref{froisconst}), in 
which case the fit can be extended down to $\sqrt{s}_{min}=5 GeV$. 

In light of our analysis, we can give two possible explanations. One is 
that the later fit is a coincidence, which would be supported by the 
fact that one extends the fit down in energies, not up, which seems 
counterintuitive. Also, the reaction dependent constant term  is maybe 
less motivated and indicative of a coincidence due to 
having many parameters in the fit (note though that both the ``soft Pomeron''
and the later fit have the same number of parameters). Finally, the 
coefficient B of $ln^2(s/s_0)$ is only about $\sim 0.31 mb\ll \pi/m_{\pi}^2
=60 mb$, with constant term $Z_{AB}\sim 18-65 mb$, compared to $X_{AB}\sim 
10-35mb$ for the soft Pomeron fit. Even though B should be $<\pi/m_{\pi}^2$,
Heisenberg's model gave saturation also for the coefficient, so we would expect
B at least to be close to $60mb$, while one finds that instead
$Z_{AB}$ reaches up to $65mb>60 mb$.
On the other hand, for the soft Pomeron one has $X_{AB}\sim 10-35 mb$, so 
maybe only when $X_{AB}(s/s_0)^{\epsilon}$ reaches $\sim 60 mb$ we will 
turn over to the Heisenberg (maximal Froissart) behaviour. 

Another possibility is that in our analysis, due to the fact that 
$M_L=m_{\pi}$ is so small compared to $\Lambda_{QCD}$,  we have 
$\hat{E}_F'$ (the onset of maximal Froissart behaviour in $m_{\pi}$) is lower 
or of the order of $\hat{E}_R$, even though as we explained, $\hat{E}_F
>\hat{E}_R$ (the maximal Froissart behaviour in $M_1$ should onset after 
the $s^{1/11}$ behaviour. So one creates black holes in almost flat space, but
then the bending of the IR brane catches up before the black holes 
can feel the curvature of AdS. 
Thus it could be that both the ``soft Pomeron''
behaviour $s^{1/11}$ and the maximal Froissart behaviour in $m_{\pi}$ coexist, 
and so the real behaviour would be 
\bea
&&\sigma_{AB}- Y_{1AB}(s_1/s)^{\eta_1}+Y_{2AB}(s_1/s)^{\eta_2}= 
X_{AB}(s/s_0)^{\epsilon} +B log^2(s/s_0) 
\nonumber\\&&
\sigma_{\bar{A}B}- Y_{1AB}(s_1/s)^{\eta_1}-Y_{2AB}(s_1/s)^{\eta_2}= 
X_{AB}(s/s_0)^{\epsilon} +B log^2(s/s_0)
\label{combo}
\eea

It would be hard to imagine though that one will then still get the 
same remarkable agreement with $\epsilon=0.0933\pm 0.0024$ onsetting 
at $9GeV$.

In either case, we would suggest that there is need for a further 
experimental work, as most of the data is situated around 10 GeV, and 
further up around 1 TeV, and not much in between. 

If indeed one observes the soft Pomeron behaviour, either by itself as in 
(\ref{softp}) or together with the maximal 
Froissart behaviour as in (\ref{combo}),
we have an experimental test of string theory, literally counting the 
extra dimensions, since as we have seen one has $\sigma \sim s^{1/n}$ and 
n comes from the behaviour of the Laplacean on $AdS_{d+1}\times X_{\bar{d}}
$ as $1/r^n= 1/r^{2(d-1)+\bar{d}}$.  

Of course it would be nice to find the real gravity dual to QCD, so that
one can compute precisely the scales $\Lambda_{QCD},\hat{M}_P,
\hat{E}_R, \hat{E}_F, \hat{E}_F'$ 
in particular, and distinguish between the various possibilities. 
It would also be nice to have a scattering description for the pion Heisenberg
(maximal Froissart) behaviour, dual to brane bending. 
Short of that, we have presented a consistent analysis, and shown that the 
soft Pomeron behaviour can count the extra dimensions of string theory, 
and it corresponds in QCD to creation of an effective field theory 
``soliton''-like structure that then decays, mapped to black hole production
in the gravity dual.

{\bf Acknowledgements} I would like to thank Kyungsik Kang for many 
discussions and for pointing out ref. \cite{compas,compete,pdgold,pdg}
and explaining the experimental evidence to me, and thank Micha 
Berkooz and Peter Freund for discussions. 
This research was  supported in part by DOE
grant DE-FE0291ER40688-Task A.

\newpage

{\Large\bf{Appendix A. Solution for $\Phi$ and trapped surfaces.}}

\renewcommand{\theequation}{A.\arabic{equation}}
\setcounter{equation}{0}

\vspace{1cm}

In this Appendix we solve for the A-S solution in the $AdS_5\times X_5$ 
background (with $X_5$ very large, almost flat), and calculate the trapped 
surface that forms in the collision of two such shockwaves. 

The equation we need to solve is the Poisson equation in the background,
\be
\Delta \Phi = -16\pi G_{d+\bar{d}+1} p \delta ^{d-2} (x^i) \delta (y-y_0)
\delta ^{\bar{d}}(z_i)
\ee
where $\Delta $ is the laplacean for massless particles propagating in the 
$AdS_{d+1}\times X_{\bar{d}}$ background, i.e.
\be
\Delta = \nabla_x^2 + e^{-2y/R} (\partial_y^2-\frac{d}{R}\partial_y)
+ e^{-2y/R}\partial_{z^i}^2
\ee
Here $x^i$ are d-2 flat transverse coordinates on the brane, y is the 
AdS radial coordinate and $z^i$ are $\bar{d}$ 
(almost) flat coordinates on $X_{\bar{d}}$. 

A Fourier transform 
\be
\phi(q, \bar{q},y)
= \int d^{d-2}x e^{-iq\cdot x}\int d^{\bar{d}}z e^{-i\bar{q}\cdot z} 
\Phi(x, y, z)
\ee
turns the equation into (outside $y=y_0=0$. We choose $y_0=0$ for simplicity,
as it can be simply reintroduced by rescaling)
\be
\phi '' (q, \bar{q}, y)-\frac{d}{R} \phi ' (q, \bar{q}, y)- (q^2 e^{2y/R}
+\bar{q}^2) \phi (q, \bar{q}, y)=0
\ee
We see that for $y/R\ll 1$, the equation is the same as the equation for 
$AdS_{d+1}$ found in \cite{kn}, in variables $Q^2= q^2+\bar{q}^2$. Therefore
the solution for $y/R\ll 1$ is 
\bea
&&\Phi (r, y)= \frac{4 G_{d+\bar{d}+1}p e^{\frac{dy}{2R}}}{2\pi} 
\int \frac{d^{d-2}q}{(2\pi)^{d-2}}e^{i\vec{q}\vec{x}}\int \frac{d^{\bar{d}}\bar
{q}}{(2\pi)^{\bar{d}}}e^{i\bar{q}\cdot z}K_{d/2} (e^{y/R}RQ)I_{d/2}(RQ)
\nonumber\\
&&= \frac{4 G_{D+1}p e^{\frac{dy}{2R}}}{(2\pi)^{D+1}}
\frac{1}{r^{\frac{D-4}{2}}}\int _0^{\infty} dQ Q^{\frac{D-2}{2}}J_{\frac{
D-4}{2}}(Qr) K_{d/2}(e^{y/R}RQ)I_{d/2}(RQ)\nonumber\\&&
=\frac{\bar{C}e^{\frac{dy}{2R}}}{r^{D-2}}\int_0^{\infty} dw 
w^{\frac{D-2}{2}}J_{\frac{D-4}{2}}(w)K_{d/2}(e^{y/R}\frac{Rw}{r})
I_{d/2}(\frac{Rw}{r})
\label{asbgr}
\eea
where $r^2= \vec{x}^2+\vec{z}^2$,
$D=d+\bar{d}$ and $\bar{C}= 8G_{D+1}l p/(2\pi)^{(D-4)/2}$, and the 
solution is valid for $y>0$. If $y<0$, we should exchange $K_{d/2}$ and 
$I_{d/2}$. We are interested in the behaviour of $\Phi $ at large $r/R$, thus 
for small argument of $K_{d/2}$ and $I_{d/2}$.

When we expand $K_{d/2}(aw)I_{d/2}(bw)$ at small w we obtain mostly
 $w^{2n}$ terms, but they  will 
give a zero result after integration, since 
\be
\int_0^{\infty} dw w^{D/2-1+2n} J_{D/2-2}(w)=0
\ee
The first nonzero result comes when we get a $w^{n}log(w)$ term. We can 
check that the first such term in the expansion of $K_{d/2}(aw)I_{d/2}(bw)$
is $\sim f a^{d/2}b^{d/2}w^{d}log (w)$, (f=numerical constant)
and this gives a nonzero integral.

Thus the behaviour of $\Phi$ is 
\be
\Phi (r\gg R, y=0) = \frac{K_1 R_s R^{D+d-2}}
{r^{D+d-2}}= \frac{const.}{r^{2(d-1)+\bar{d}}
}
\ee
where
\be
K_1= \frac{f}{(2\pi)^{(D-4)/2}} \int_0^{\infty} dw w^{d+\frac{D-2}{2}}
J_{\frac{D-4}{2}}(w)log (w)
\ee

In the following we restrict to the physical case $d=4, \bar{d}=5$,
but present some formulas at general $d,\bar{d}$. The 
calculation mirrors exactly the one in Appendix A of \cite{kntwo}.

We have 
\be
I_2(bx)K_2(ax)= \frac{b^2}{4a^2}+ct. x^2 +ct. x^4- \frac{1}{64}a^2 b^2 x^4 
log (x)+o(x^5)
\ee 
and using
\be
I= \int_0^{\infty} dw w^{4+7/2} J_{5/2}(w) log (w)=-3780 \sqrt{2\pi}
\ee
we get 
\be
\Phi (r\gg R, y=0)\simeq (-\frac{I}{64})\frac{\bar{C}R^4}{r^{11}}=
\frac{945\sqrt{2\pi}}{16}
 \frac{\bar{C}R^4}{r^{11}}
\ee
and 
\be
\Phi (r\gg R, y/R \ll 1) \simeq \frac{\bar{C} e^{2y/R}}{r^7}
\int_0^{\infty} dw w^{7/2} J_{5/2}(w) K_2 (e^{y/R}\frac{Rw}{r})I_2(
\frac{Rw}{r})
\ee
Using $ wK_{\nu}'(w)+\nu K_{\nu}(w)=-w K_{\nu -1}(w)$, we get 
\be
\partial_{y}\Phi= -\frac{\bar{C}e^{2y/R}}{Rr^7}
\int_0^{\infty} dw w^{7/2} J_{5/2}(w) (e^{y/R}\frac{Rw}{r})K_1
(e^{y/R}\frac{Rw}{r})I_2(\frac{Rw}{r})
\ee
Then
\be
\partial_y \Phi|_{y=0}= -\frac{\bar{C}}{r^8}\int _0^{\infty} dw w^{9/2}
J_{5/2}(w) K_1 (\frac{Rw}{r})I_2(\frac{Rw}{r})
\ee
and 
\be
\partial^2_y\Phi|_{y=0}= \frac{2}{R}\partial_y \Phi|_{y=0} +\frac{\bar{C}}{
r^9}\int_0^{\infty} dw w^{11/2} J_{5/2} (w) K_0(\frac{Rw}{r}) I_2(\frac{Rw}{r})
\ee
Expanding the Bessel functions at small argument
\bea
&&K_0(x)I_2(x)\simeq (-\frac{\gamma}{8}+\frac{log(2)}{8}-\frac{log(x)}{8})x^2+
o(x^4)\nonumber\\
&& K_1(x)I_2(x)\simeq \frac{x}{8}+(-\frac{1}{48}+\frac{\gamma}{16}
-\frac{log(2)}{16}+\frac{log(x)}{16})x^3+o(x^5)
\eea
we get 
\bea
&& \partial_y\Phi|_{y=0}\simeq(-\frac{I}{16})\frac{\bar{C}R^3}{r^{11}}=\frac{4
\Phi|_{y=0}}{R}\nonumber\\
&&\partial_y^2\Phi|_{y=0}\simeq
\frac{8}{R^2}\Phi|_{y=0}-\frac{I}{8} \frac{\bar{C}
R^2}{r^{11}}= \frac{16 \Phi|_{y=0}}{R^2}
\eea
The trapped surface condition is now
\be
(\partial_i\Psi)^2+e^{-2y/R}(\partial_y\Psi)^2+e^{-2y/R}(\partial_{\mu}
\Psi)^2=4;\;\; \Psi=\Phi+\zeta
\ee
where $\mu=1,5$ are indices for the $z^{\mu}$
coordinates on $X_5$. We work at z=0 (fixed position 
in the extra dimensions), thus $(\partial_{\mu}\Phi)^2|_{z=0}=0$. 

The above condition is matched against $\Psi= C$=const. to find $\zeta$
perturbatively, $\zeta= \zeta_0(r)+\zeta_1(r)+\zeta_2(r)y^2/2$, with 
$\zeta_0=0$ for consistency. Then 
\be
\Psi= f+ ay +\frac{y^2}{2}g+...
\ee
where
\bea
&&f=\Phi|_{y=0}=(-I/64)\frac{\bar{C}R^4}{r^{11}}= \frac{K_1 R_s R^n}{r^n}
\nonumber\\&&
a= \partial_y\Phi|_{y=0}+\zeta_1=\frac{4}{R}f +\zeta_1\nonumber\\
&&g=\partial_y^2 \Phi|_{y=0}+\frac{d}{R}\zeta_1= \frac{16f}{R^2}
+\frac{4\zeta_1}{R}
\eea
If a is nonzero, one then has to match 
\bea
&&4=f'^2 +a^2+y(2a'f'-\frac{2a^2}{R}+2ag)+...\nonumber\\&&
{\rm vs.}\;\;C=f+ay
\eea
If $\zeta_1=0$ one obtains
\bea
&& 1=(\frac{2f}{R})^2[1+\frac{6y}{R}+(\frac{nR}{4r})^2+...]\nonumber\\
&&{\rm vs.}\;\; C^2=f^2[1+\frac{8y}{R}]
\eea
which has no solution, thus one needs to take a nonzero $\zeta_1$. 
If one takes $\zeta_1$ and $a$ of the same order, giving $a=-\alpha f/R$,
we get 
\bea
&& 1=(\frac{\alpha f}{2R})^2 [1+\frac{6y}{R}+(\frac{nR}{\alpha r})^2+...]
\nonumber\\&& {\rm vs.} C^2=f^2[1-\frac{2\alpha y}{R}+...]
\eea
which has the solution $\alpha=-3$. Thus we have a trapped surface with 
\be
a=\frac{3f}{R}
\ee
which gives the condition for the size of the trapped surface
\be
\frac{3f}{2R}=1
\ee
and at nozero impact parameter of the two colliding A-S shockwaves we get 
approximately
\be
(\frac{3f}{2R})^2 (1-\frac{b^2}{2r^2})=1
\ee
With the defintions
\be
f=\frac{K_1 R_s R^n}{r^n};\;\; (\frac{3K_1 R_s R^n}{2R})^2=a;\;\;\; r^2=x
\ee
one has to solve the equation
\be
g(x)= x^{n+1}-ax +\frac{ab^2}{2}=0
\ee
The maximum b for which there is a solution is found from $g'(x_0)=0, g(x_0)=0
$, giving
\be
b^2_{max}= \frac{2n}{n+1}(\frac{a}{n+1})^{\frac{1}{n}}
\Rightarrow b_{max}= \sqrt{\frac{2n}{n+1}}(n+1)^{-1/n}R (\frac{3K_1R_s}{2R})
^{\frac{1}{n}}
\ee
As in \cite{kntwo} one finds an extra trapped surface that would be there 
in the absence of AdS, and that surface is smaller. The same physical argument,
that the large warping of AdS is expected to create a larger black hole, 
applies. Therefore, one will take the above trapped surface 
solution as describing best the horizon of the formed black hole.

\end{document}